%% file: main.tex
\documentclass[sigconf,nonacm]{acmart}
\usepackage{etex}
\usepackage[utf8]{inputenc}
\usepackage[ruled,vlined]{algorithm2e}
\usepackage[T1]{fontenc}
\usepackage{microtype}
\usepackage{multirow}
\usepackage{enumitem}
\usepackage{colortbl}
\usepackage{pifont}
\usepackage{xspace}
\usepackage{url}
\usepackage{rotating}
\usepackage[TABBOTCAP]{subfigure}
\usepackage{listings}
\usepackage{algorithmic}
\usepackage{textcomp}
\usepackage{xcolor}
\usepackage{fontawesome}
\usepackage{calc}
\usepackage[table]{xcolor}

\input{macro}

\begin{document}
\settopmatter{authorsperrow=2}
\title[On the Impact of Code Comments for Automated Bug-Fixing: An Empirical Study]{On the Impact of Code Comments for Automated Bug-Fixing:\\An Empirical Study}

\author{Antonio Vitale}
\email{antonio.vitale@polito.it}
\affiliation{%
  \institution{Politecnico di Torino, University of Molise}
  \country{Italy}
}

\author{Emanuela Guglielmi}
\email{emanuela.guglielmi@unimol.it}
\affiliation{%
  \institution{University of Molise}
  \country{Italy}
}

\author{Simone Scalabrino}
\email{simone.scalabrino@unimol.it}
\affiliation{%
  \institution{University of Molise}
  \country{Italy}
}

\author{Rocco Oliveto}
\email{rocco.oliveto@unimol.it}
\affiliation{%
  \institution{University of Molise}
  \country{Italy}
}

\renewcommand{\shortauthors}{Vitale \etal}

\begin{abstract}
Large Language Models (LLMs) are increasingly relevant in Software Engineering research and practice, with Automated Bug Fixing (ABF) being one of their key applications.
ABF involves transforming a \textit{buggy} method into its \textit{fixed} equivalent.
A common preprocessing step in ABF involves removing comments from code prior to training.
However, we hypothesize that comments may play a critical role in fixing certain types of bugs by providing valuable design and implementation insights.
In this study, we investigate how the presence or absence of comments, both during training and at inference time, impacts the bug-fixing capabilities of LLMs.
We conduct an empirical evaluation comparing two model families, each evaluated under all combinations of training and inference conditions (\textit{with} and \textit{without} comments), and thereby revisiting the common practice of removing comments during training.
To address the limited availability of comments in state-of-the-art datasets, we use an LLM to automatically generate comments for methods lacking them.
Our findings show that comments improve ABF accuracy by up to threefold when present in both phases, while training with comments does not degrade performance when instances lack them.
Additionally, an interpretability analysis identifies that comments detailing method implementation are particularly effective in aiding LLMs to fix bugs accurately.
\end{abstract}

\ccsdesc[500]{Software and its engineering~Software development methods}
\ccsdesc[500]{Software and its engineering~Empirical software validation}
\ccsdesc[500]{Computing methodologies~Machine learning}

\maketitle

\input{introduction}
\input{related-work}
\input{theory}

\input{study}
\input{results}
\input{discussion}
\input{threats}
\input{conclusion}

\begin{acks}
This publication is part of the project PNRR-NGEU which has received funding from the MUR – DM 118/2023. This work has been partially supported by the European Union - NextGenerationEU through the Italian Ministry of University and Research, Projects PRIN 2022 “QualAI: Continuous Quality Improvement of AI-based Systems”, grant n. 2022B3BP5S, CUP: H53D23003510006.
\end{acks}

\bibliographystyle{ACM-Reference-Format}
\bibliography{main}
\end{document}

%% file: macro.tex
\lstdefinestyle{mystyle}{
	backgroundcolor=\color{backcolour},
	commentstyle=\color{codegreen},
	keywordstyle=\color{codepurple},
	numberstyle=\tiny\color{codegray},
	stringstyle=\color{codemagenta},
	language=Java,
	breakatwhitespace=false,
	breaklines=true,
	keepspaces=true,
	numbers=none,
	numbersep=5pt,
	showspaces=false,
	showstringspaces=false,
	showtabs=false,
	tabsize=2,
	frame=tb,
	framerule=0pt,
	basicstyle=\fontsize{5.5}{5.5}\fontfamily{\ttdefault}\selectfont
}
\lstdefinestyle{mystyleresult}{
	backgroundcolor=\color{backcolour},
	commentstyle=\color{codegreen},
	keywordstyle=\color{codeoutput},
	numberstyle=\tiny\color{codegray},
	stringstyle=\color{red},
	language=Java,
	breakatwhitespace=false,
	breaklines=true,
	keepspaces=true,
	numbers=none,
	numbersep=5pt,
	showspaces=false,
	showstringspaces=false,
	showtabs=false,
	tabsize=2,
	frame=tb,
	framerule=0pt,
	basicstyle=\color{codeoutput}\fontsize{5.5}{5.5}\fontfamily{\ttdefault}\selectfont
}

\newcommand{\ie}{\emph{i.e.,}\xspace}
\newcommand{\eg}{\emph{e.g.,}\xspace}
\newcommand{\etc}{etc.\xspace}
\newcommand{\etal}{\emph{et~al.}\xspace}
\newcommand{\secref}[1]{Section~\ref{#1}\xspace}
\newcommand{\figref}[1]{Fig.~\ref{#1}\xspace}

\newcommand{\tabref}[1]{Table~\ref{#1}\xspace}

\newcommand{\RQ}[1]{RQ$_{#1}$\xspace}

\definecolor{Gray}{gray}{0.9}
\definecolor{codegreen}{rgb}{0,0.6,0}
\definecolor{codegray}{rgb}{0.73,0.38,0.06}
\definecolor{codepurple}{rgb}{0.70,0.27,0}
\definecolor{codemagenta}{rgb}{0.74,0.09,0.42}
\definecolor{codeoutput}{rgb}{0.5,0,0}
\definecolor{backcolour}{rgb}{0.96,0.96,0.96}

\newboolean{showcomments}
\setboolean{showcomments}{true}
\ifthenelse{\boolean{showcomments}}{
  \marginparwidth=1.35cm
  \newcommand{\nb}[2]{\todo[size=\tiny, color=green!40, nolist, fancyline]{\textbf{\textsc{#1}}. #2}}
}{
  \newcommand{\nb}[2]{}
}

\definecolor{gray50}{gray}{.5}
\definecolor{gray40}{gray}{.6}
\definecolor{gray30}{gray}{.7}
\definecolor{gray20}{gray}{.8}
\definecolor{gray10}{gray}{.9}
\definecolor{gray05}{gray}{.95}

\newlength\Linewidth
\def\findlength{\setlength\Linewidth\linewidth
  \addtolength\Linewidth{-4\fboxrule}
  \addtolength\Linewidth{-3\fboxsep}
}
\newenvironment{examplebox}{\par\addvspace{4pt}\begingroup
  \setlength{\fboxsep}{5pt}\findlength
  \hspace{-0.4cm}
  \setbox0=\vbox\bgroup\noindent
  \hsize=0.95\linewidth
  \begin{minipage}{0.95\linewidth}\normalsize}
  {\end{minipage}\egroup
  \textcolor{gray20}{\fboxsep1.5pt\fbox
    {\fboxsep5pt\colorbox{gray05}{\normalcolor\box0}}}
  \endgroup\par\noindent
  \normalcolor\ignorespacesafterend}

\newenvironment{findbox}{
  \par%
  \begingroup
  \setlength{\fboxsep}{6pt}%
  \hspace{-0.4cm}%
  \setbox0=\vbox\bgroup\noindent
  \hsize=0.95\linewidth
  \begin{minipage}{0.95\linewidth}\normalsize
}{%
  \end{minipage}\egroup
  \begingroup
    \fboxrule=0.6pt
    \fcolorbox{gray!80}{gray!10}{\box0}%
  \endgroup
  \endgroup\par\noindent
  \normalcolor\ignorespacesafterend
}

\newenvironment{rqbox}{
  \par%
  \begingroup
  \setlength{\fboxsep}{6pt}%
  \hspace{-0.4cm}%
  \setbox0=\vbox\bgroup\noindent
  \hsize=0.95\linewidth
  \begin{minipage}{0.95\linewidth}\normalsize
}{%
  \end{minipage}\egroup
  \begingroup
    \fboxrule=0.6pt
    \fcolorbox{black}{gray!20}{\box0}%
  \endgroup
  \endgroup\par\noindent
  \normalcolor\ignorespacesafterend
}

%% file: introduction.tex
\section{Introduction} \label{sec:intro}

Large Language Models (LLMs), are increasingly becoming indispensable tools for assisting software developers in their daily tasks. These models enable the automation of a wide range of activities. Furthermore, developers can easily leverage advanced LLM-powered tools, such as GitHub Copilot, which are seamlessly integrated into modern Integrated Development Environments (IDEs) to enhance productivity and streamline workflows.

Among these applications, Automated Bug Fixing (ABF) has emerged as one of the most researched and impactful coding tasks \cite{tufano2019empirical,mastropaolo2021studying,berabi2021tfix,lutellier2020coconut,chen2019sequencer,ye2022neural,jiang2021cure}. ABF involves identifying and resolving bugs within a method (\eg a scenario where a test case fails). While the concept may seem straightforward, ABF holds significant value for developers, as bug fixing is notoriously challenging and often demands considerable effort \cite{eladawy2024automated,o2017debugging}. By alleviating this burden, ABF has the potential to substantially enhance developer efficiency and reduce the time spent on error resolution.

Early research on using deep learning (DL) and LLMs for ABF established the foundation for the field. Tufano \etal \cite{tufano2019empirical} presented one of the first DL-based approaches to automatically fix bugs, quickly followed by several others \cite{chen2019sequencer,lutellier2020coconut,ye2022neural}. A key contribution of the work by Tufano \etal \cite{tufano2019empirical} was the release of a benchmark dataset, which remains a standard for training and evaluating ABF models \cite{wang2023rap,mastropaolo2021studying,chakraborty2021multi,drain2021generating,wang2021codet5,ahmad2021unified,guo2020graphcodebert,huang2023empirical,huang2025comprehensive}. Such a dataset, mined from GitHub, contains couples of buggy Java methods and their respective fixed versions. Given the limitations of the specific ML model they adopted (Recurrent Neural Networks --- RNNs), they needed to pre-process the source code of the methods so that identifiers are abstracted (\ie they appear like \texttt{IDi}, where \texttt{i} is an incremental number) and comments are removed.

\begin{figure}[t]
    \centering
    \includegraphics[width=0.9\linewidth]{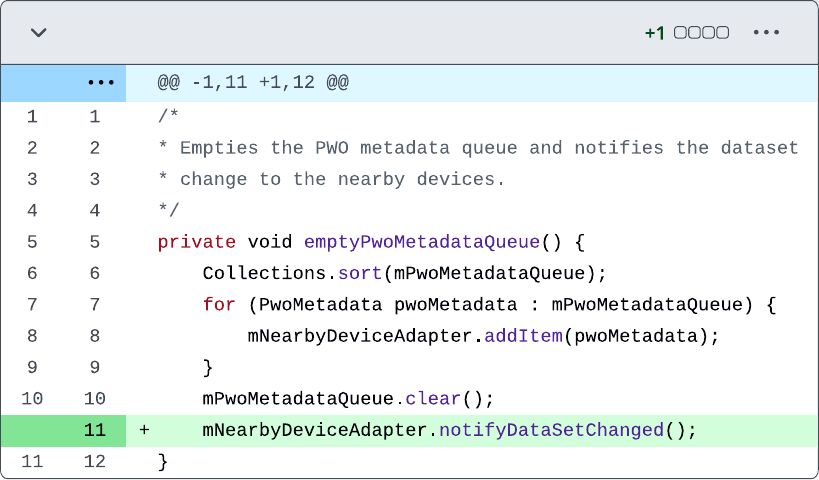}
    \caption{Example of a buggy method and its fixed version (the green line is the fix, which was missing in the buggy version).}
    \label{fig:intro:example}
\end{figure}

However, code comments may not only be important \cite{vitale2024catalog}, but often essential for fixing bugs. Comments, indeed, might contain precious information about the requirements, which are sometimes necessary to understand what is wrong with the code and, thus to fix it. Let us consider the example in \figref{fig:intro:example}. The buggy code lacks \texttt{mNearbyDeviceAdapter.notifyDataSetChanged()}.
Such an instruction notifies the dataset change to the nearby devices. Without reading the comment and understanding this requirement, neither a human developer nor a model could identify or fix the bug.

Unlike the RNNs employed by Tufano \etal \cite{tufano2019empirical}, modern LLMs are well-equipped to process code comments. Despite this capability, prior research has overlooked the potential value of incorporating comments—a potentially critical source of information—for Automated Bug Fixing (ABF).

In this work, we investigate how the presence or absence of code comments, both during training and at inference time, impacts on the bug-fixing capabilities of Large Language Models (LLMs).
First, we build a new dataset for ABF based on the one introduced by Tufano \etal \cite{tufano2019empirical}. To do this, we consider the same repositories and commits, but we re-extract the instances by considering the raw source code (\ie we do not abstract identifiers and keep the original comments). Differently from the previous work, which only considered \textit{small}- ($\leq$ 50 tokens) and \textit{medium}-sized ($\leq$ 100 tokens) methods in two distinct datasets, we merge them in a single dataset, in which we also include \textit{big} methods ($\leq$ 512 tokens).

We noticed, however, that only a minority of the code instances in our dataset contained comments (< 20\%) and that many of them are low quality (\eg \texttt{``// Argh!!! This method is WRONG !''}). For this reason, we use an automated procedure to generate higher-quality comments needed in our experiment. Considering the taxonomy of code comments introduced by Zhai \etal \cite{zhai2020cpc}, we use an LLM to generate five comments, documenting \textbf{what} the method does, \textbf{how} it does it, the intended \textbf{usage} of the method, the reason \textbf{why} the method is there, and \textbf{properties} of the method (\eg pre-conditions and post-conditions). We do this on the fixed version rather than on the buggy one to simulate the ideal scenario in which developers properly document the methods they develop.

To evaluate our hypothesis that code comments play a crucial role in Automated Bug Fixing (ABF), we designed a controlled experiment manipulating the presence of comments both during training and inference. We fine-tuned two types of models—CodeT5+ \cite{wang2023codet5+} and DeepSeek-Coder \cite{guo2024deepseek}—under all combinations of these conditions (\textit{with} vs. \textit{without} comments). For the training phase, comments were either removed from or automatically generated for each instance, while at inference time the models were tested on inputs with or without comments accordingly.

When comparing the four training-inference combinations, we found that models benefit from comments in both phases. Comments provided only at inference time already lead to noticeable improvements, while training with comments does not harm performance on instances lacking comments. The greatest improvements occur when comments are present in both phases, increasing accuracy by up to threefold. These results demonstrate that code comments can substantially enhance models performance, underscoring their importance in ABF tasks.

To gain deeper insights into the impact of the five categories of comments on model performance, we conducted an interpretability study based on SHAP \cite{lundberg2017unified}, employing the GradientExplainer method.
This method assigns an importance score to each input token with respect to every token generated during the inference of a single generation. Our analysis revealed that the most influential category of comments were those documenting \textbf{how} the method achieves its goal.

Interestingly, the importance of other comment categories varied between the two models. For one model, \ie CodeT5+, comments describing \textbf{what} the method does were the least important, likely because they provide high-level information that is often redundant (\eg already conveyed by the method name). For the other model, \ie DeepSeek-Coder, the lowest importance was attributed to comments describing the method's \textbf{usage} and its \textbf{properties}.

In summary, our findings reinforce a recurring theme in Software Engineering research: \emph{Developers should document their code not only for themselves and future collaborators but also to enable LLMs to provide better assistance}.

All code and data used in our study is publicly available \cite{replicationpackage}.

%% file: related-work.tex
\section{Background and Related Work}
\label{sec:related}
We first present some background on Machine Learning (ML) for coding tasks, with specific focus on the Automated Bug Fixing (ABF) task. Then, we present some related works.

\subsection{ML for Coding Tasks}
Software Engineering ML tasks are typical software engineering activities that can be supported by Machine Learning.
Among these, Coding ML Tasks (or simply Coding Tasks) concern source code.
They may involve code classification (\eg vulnerability detection), regression (\eg bug-fix time prediction), or code generation (\eg automated bug fixing).
We focus on the latter, which includes code summarization \cite{leclair2019recommendations, ahmad2020transformer, haque2020improved, mastropaolo2024towards}, code completion \cite{svyatkovskiy2020intellicode, ciniselli2021empirical}, and automated bug fixing \cite{tufano2019empirical, chen2019sequencer}.
While several approaches tried to represent source code as trees (\eg AST) or graphs (\eg CFG) to devise ML-approaches able to perform specific tasks, most models simply represent source code as text. Thus, neural networks used in Natural Language Processing (NLP) tasks found their way into the SE community.
The first approaches defined to tackle coding tasks were based on Recurrent Neural Networks (RNNs).
RNNs are designed to process sequential data through recurrent connections that allow information to persist across sequence steps.
Their basic input unit is the \textit{token}, which represents a word, subword, or character, depending on the tokenizer used to convert text into token sequences.
Since RNNs handle numeric vectors, tokens are transformed into vector representations through an \textit{embedding} step.

The Transformer architecture \cite{vaswani2017attention} constituted a revolution in this field. Such an architecture has been adopted both for NLP \cite{raffel2020exploring} and coding tasks \cite{mastropaolo2021studying, mastropaolo2022usingt5, wang2021codet5}. 
The Transformer architecture relies on a self-attention mechanism to capture long-range dependencies among tokens. Unlike RNNs, which process tokens sequentially, this architecture can handle longer context windows (\ie larger amounts of tokens) efficiently given its parallel tokens processing.

While the definition of \textit{Large Language Model} (LLMs) is fuzzy, most use such a term to refer to any artificial neural network that relies on the Transformer architecture \cite{vaswani2017attention} to perform a specific task. Thus, we use such a term with this meaning from now on.

LLMs work in two steps: \textit{pre-training} and \textit{fine-tuning}. In the \textit{pre-training} step, the model is trained to perform a self-supervised task (\eg predicting
the next token in a sequence). This allows the model to learn the language(s) that it must treat. In the \textit{fine-tuning} step, instead, the model is trained to perform a specific task (\eg to transform a given input sequence to a given output sequence). For example, for the ABF task on which we focus in this paper, the input sequence represent a \textit{buggy} method, while the output sequence is the \textit{fixed} version of the same method.
Since 2021, several LLMs specialized in source code have emerged \cite{mastropaolo2021studying, mastropaolo2022usingt5, wang2021codet5, wang2023codet5+}.
Among them, CodeT5+ \cite{wang2023codet5+} extends the T5 architecture \cite{raffel2020exploring}.
It is pre-trained on nine programming languages using multiple code-oriented objectives, including \textit{span denoising}, and \textit{causal LM}, and is available in sizes from 220M to 16B parameters.
DeepSeek-Coder \cite{guo2024deepseek}, instead, is trained from scratch on 2T tokens across 87 languages, building on the Llama architecture \cite{touvron2023llama}.
It employs \textit{next-token prediction} and \textit{fill-in-the-middle} pre-training objectives, and is available in base and instruction-tuned variants ranging from 1.3B to 33B parameters.

\subsection{Automated Bug Fixing}
One of the first studies regarding the use of DL to fix bugs is the one by Tufano \etal \cite{tufano2019empirical}. The authors use a Neural-Machine-Translation (NMT) approach to address ABF. They found that it was able to fix between 9\% and 50\% of the bugs from their dataset.

Lutellier \etal \cite{lutellier2020coconut} proposed \textsc{CoCoNuT}, a program repair technique combining ensemble learning and context-aware NMT. The method uses convolutional neural networks (CNNs) to encode buggy code and context independently, enhancing accuracy and capturing a wide range of strategies. \textsc{CoCoNuT} was observed to outperform 27 strategies in four languages, fixing 509 bugs (309 of which were unique).

Previous work mostly rely on word-level tokenizers that split the input/output text into tokens based on words defined from a given vocabulary. Therefore, they inherently suffer from the out-of-vocabulary (OOV) problem: If a word not belonging to the vocabulary appears in the test data, it can not be handled and it is often replaced with a special token (\eg \texttt{<UNK>}). Also, they mostly relied on Recurrent Neural Networks, which are not capable of considering a large context while predicting tokens.

The introduction of Transformers and advanced tokenizers like Byte Pair Encoding (BPE) \cite{kudo2018sentencepiece} allowed to overcome such limitations.
Mastropaolo \etal \cite{mastropaolo2021studying, mastropaolo2022using} relied on the T5 architecture to tackle several coding tasks, including ABF. Their results show that such an approach achieves better results than the approach by Tufano \etal \cite{tufano2019empirical} on their two datasets.

Jiang \etal \cite{jiang2023impact} studied several \textit{Code Language Models} -- LLMs for coding tasks -- \ie PLBART \cite{ahmad2021unified}, CodeT5 \cite{wang2021codet5}, CodeGen \cite{nijkamp2022codegen}, and InCoder \cite{fried2022incoder} in the bug-fixing context. The authors showed that the best CLM, without fine-tuning, fixes 72\% more bugs than the state-of-the-art deep learning-based techniques, while fine-tuning enables CLMs to fix 46\%–164\% more bugs than previous techniques.

\subsection{Data Quality in Coding Tasks}
A few studies analyzed the impact of training data quality on the resulting model.
Liu \etal \cite{liu2022learning} addressed the method-name prediction task enriching the instances with (i) documentation, (ii) local context, and (iii) project-specific context. This allowed them to train a model which outperformed state-of-the-art approaches by a large margin.
Prenner \etal \cite{prenner2024out} studied the importance of local context in instances for automated program repair models. Given different measures of context (\eg lines before the bug), the authors trained different Transformer models, finding that the repair success increases when the local context is bigger.
The work most closely-related to ours is the one by Van Dam \etal \cite{van2023enriching}. The authors show that enriching the code instances for the code completion task with their original comments results in higher effectiveness.
Differently from Van Dam \etal, we (i) focus on the ABF task, and (ii) do not rely on the original comments by the authors but on possibly higher-quality and more complete comments, automatically generated with GPT-3.5.

%% file: theory.tex
\section{Theory: Why and What Comments Matter}
\label{sec:theory}

In this section, we present our theory on the significance of comments for ABF through LLMs and outline the types of informative content in the comments that we expect could play a crucial role.

\subsection{Why Comments Matter for ABF with LLMs}
LLMs are, at their core, language models: Their goal is to predict the most likely token $t_{n+1}$ given a \textit{context} $\chi$, which is a sequence of tokens $\langle t_1, \dots, t_n \rangle$. 
The fine-tuning step, which teaches the model how to perform a specific task, is performed through a dataset $\Phi$, in which each instance is represented as a pair $\langle \mathit{input}, \mathit{output} \rangle$, where both elements are token sequences.
Let us consider the ABF task, \ie the task of transforming a given \textit{buggy} method $m_b$ into a non-buggy (or fixed) method $m_f$. In the fine-tuning step, the model is trained to generate, one at a time, the tokens in $m_f$ (\textit{output}) given the ones in $m_b$ (\textit{input}). LLMs are auto-regressive models \cite{vaswani2017attention}: For each token to generate, the context is constituted by the union of the \textit{input} sequence ($m_b$ for ABF) and the previously generated tokens, until a special token is generated to indicate the end of the \textit{output} sequence. Formally, the context that is adopted to generate the token $t_k$ can be recursively expressed as $\chi_{t_k} = \chi_{t_{k - 1}} \cup G(\chi_{t_{k - 1}})$, where $\chi_{t_0}$ is simply the input sequence (buggy method $m_f$ in ABF) and $G$ indicate the inference procedure that predicts the next token given a context.

Consider two identical methods, $m_a$ and $m_b$, both consisting of a single line of Python code: \texttt{return sorted(list)}. The first method, $m_a$, is intended to sort a list of strings in a case-sensitive manner and is therefore bug-free. In contrast, $m_b$ is meant to sort a list of strings in a case-insensitive manner, making it buggy. This example illustrates how the same source code can be simultaneously \textit{bug-free} in one context and \textit{buggy} in another.
Now, let us introduce two models: $M_{\Phi_b}$ and $M_{\Phi_c}$. The model $M_{\Phi_b}$ is fine-tuned on a dataset $\Phi_{b}$ that contains only code without comments, while $M_{\Phi_c}$ is fine-tuned on $\Phi_{c}$, a dataset enriched with comments that describe the code in each instance.
The difference between these models lies in their learning capabilities. $M_{\Phi_b}$ learns to generate code-related tokens solely based on other code-related tokens.
In essence, it focuses on repeating patterns of changes within the code itself. Conversely, $M_{\Phi_c}$ is exposed to a broader set of patterns, enabling it to learn not only code-to-code transformations but also natural language-to-code mappings, akin to what happens in the code generation task. This additional context allows $M_{\Phi_c}$ to better infer the intended behavior of the code, particularly in cases where subtle requirements—such as sorting criteria—are conveyed through comments.

\subsection{What Comments Matter}
\label{subsec:what_comments_matter}
\begin{table}[t]
\centering
\caption{Multi-intent comment taxonomy \cite{chen2021my}.}
\renewcommand{\arraystretch}{1.2}
\resizebox{0.9\linewidth}{!}{%
\begin{tabular}{l|p{4cm}|p{4cm}}
\toprule
\textbf{Category} & \textbf{Description} & \textbf{Example} \\
\midrule
What & Describes the functionality of a method & \textit{``A helper function that processes the stack.''} \\
\midrule
Why & Explains the reason why a method is provided or the design rationale of the method & \textit{``Get a copy of the map (for diagnostics)''} \\
\midrule
Usage & Describes the usage or the expected set-up of using a method & \textit{``Should be called before the object is used''} \\
\midrule
How & Describes the implementation details of a method & \textit{``Convert the byte[] to a secret key''} \\
\midrule
Property & Asserts properties of a method including pre-conditions or post-conditions of a method & \textit{``Wait until seqno is greater than or equal to the desired value or we exceed the timeout.''} \\
\midrule
Others & Unspecified or ambiguous comments & \textit{``The implementation is awesome.''} \\
\bottomrule
\end{tabular}}
\label{tab:intent_comments_categories}
\end{table}

Code comments can be aimed at documenting different aspects, such as design rationale, pre-conditions/post-conditions, and implementation details \cite{zhai2020cpc}. According to a study by Chen \etal \cite{chen2021my} comments can have six different intents, and they can be used to document \textbf{what} the method does, \textbf{why} it does it (\ie the design rationale), \textbf{how} the method is implemented, what is its intended \textbf{usage}, what \textbf{properties} characterize it (\eg pre- or post-conditions), or \textbf{other} aspects. We report in \tabref{tab:intent_comments_categories} a complete description of each category with examples.
Mu \etal \cite{mu2023developer} recently observed that over 66\% of comments in top-starred Java repositories are multi-intent, \ie they document multiple aspects of the code within a single comment. This suggests that developers often feel the need to capture various dimensions of a method behavior or purpose in their comments.
We conjecture that comments addressing all the above-mentioned intents, except for \textit{other}, can enhance the effectiveness of ABF models. Comments categorized as \textit{other} primarily represent noise and do not contribute meaningful information about the code (see the example in \tabref{tab:intent_comments_categories}). Therefore, we assume that comments with this intent play a minimal role in aiding bug fixing.

To better explain why we theorize this, let us consider the following example. There is a method named \texttt{processCSV} that reads a CSV file and processes each row to compute the mean of the numerical value contained in the $i$-th column, where $i$ is a parameter. The method is supposed to count empty cells as 0, but it throws an exception instead since it reads an empty string which is not a number. Without comments, the model might attempt to fix the method by adding a check to skip empty strings entirely. However, this would not resolve the issue but merely change the nature of the bug, as the intended behavior of counting empty cells as 0 would still not be achieved.

A \textit{property}-related comment could contain information about the post-condition (\eg ``the method returns a number'') that could help the model understand where the issue is (\ie it throws an exception). 
Comments documenting \textit{what} the method is supposed to do (\eg ``computes the mean of the values in the i-th column and counts the empty cells as 0'') and \textit{how} the method is implemented (\eg ``the method sums the values in the i-th column over the rows and returns such a value divided by the number of rows'') may help to clarify the intended functionality and find discrepancies between the code and the expected behavior. A comment documenting \textit{why} the method is necessary (\eg ``computes a robust mean of the values (for statistical calculations)'') might help to indicate the model to handle edge-cases without missing values. Finally, a comment documenting the expected \textit{usage} (\eg ``the i value should refer to a column that only contain numerical data or empty values'') could help guiding the model to avoid performing unnecessary checks (\eg if the cell contains non-numeric characters).

%% file: study.tex
\section{Empirical Study Design}
\label{sec:design}
The \emph{goal} of our study is to validate our theory, \ie to determine whether the presence of comments can improve the effectiveness of LLMs in the ABF task. Thus, our study is steered by the following Research Questions (RQs):
\begin{itemize}
\item{\RQ{1}}: \textit{Does adding comment allow to improve the bug fixing capabilities of LLMs?}
\item{\RQ{2}}: \textit{How important are the comments to the model?}
\end{itemize}

\subsection{Study Context}
The context of the study consists of buggy Java methods paired with their fixed versions. We mainly rely on the data released by Tufano \etal \cite{tufano2019empirical}.
In their work, the authors mined GitHub from March 2011 to October 2017 and collected bug-fixing commits through matching patterns (\eg ``fix bug'', ``solve error'', \etc). In total, they collected $\sim$10M of such commits.
They filtered out commits that (i) contained non-Java files, (ii) created new files, and (iii) involved changes in more than five files. Thus, they reduced the number of commits to 787,178.
They associated each \textit{fixed} version of the method to its previous version (\textit{buggy}).
After removing comments and annotations, they further abstracted the pairs by replacing identifiers and literals with placeholders, preserving frequent idioms (\ie frequent identifiers and literals).
Finally, they filtered the $\langle \mathit{bug}, \mathit{fix} \rangle$ pairs into two datasets based on the number of tokens, creating \textit{BFP}$_{small}$ (58k instances with at most 50 tokens) and \textit{BFP}$_{medium}$ (65k instances with at least 50 tokens, and at most 100 tokens).
We selected this dataset because, among those publicly available, it represents the most suitable and comprehensive option for our study. First, it provides bug-fix pairs at the \textit{method-to-method} level, offering a more realistic granularity of software changes than \textit{single-hunk} or snippet-based datasets commonly used in previous work (\eg the dataset by Zhu \etal \cite{zhu2021syntax}). Second, such instances cover a more recent and representative period (\ie 2011--2017), unlike older datasets such as those by Lutellier \etal \cite{lutellier2020coconut} (pre-2010) or Monperrus \etal \cite{monperrus2021megadiff} (circa 2015).
Finally, although the original dataset was released as two separate \textit{small} and \textit{medium} subsets, as we explain below, we merged and extended them into a single renovated version to overcome their original limitations.

\textbf{Dataset Renovation.}
Despite the before-described datasets have been extensively leveraged as benchmarks in several studies that tackled ABF \cite{wang2023rap,mastropaolo2021studying,chakraborty2021multi,drain2021generating,wang2021codet5,ahmad2021unified,guo2020graphcodebert,huang2023empirical,huang2025comprehensive}, they suffer from some limitations. Because of the limited capabilities of the models they used (\ie RNN-based) and because of the Out-of-Vocabulary (OOV) problem, the authors abstracted the identifiers of the instances and partitioned the datasets based on token count.

To overcome such limitations, we re-build the dataset from the raw source code. 
Starting with the $\sim$787k mined commits provided in the replication package by Tufano \etal \cite{tufano2019empirical}, we first extract all methods that underwent fix-related changes. Following the original methodology, we use the GumTree Spoon AST Diff tool \cite{falleri2014fine} to calculate the number of edit actions (in terms of AST transformations). We exclude methods with no edit actions — \eg those involving only changes to comments — and those requiring more than one hundred edit actions, as these are likely tangled commits involving changes beyond a simple bug fix. This step filters out 176,413 instances. Additionally, we identify and remove 87,819 duplicate instances.
We keep only instances from commits containing changes focused on a single method. We do this to avoid cases in which the fix requires edit action on more than a method. The model, indeed, will see only a single method and will be asked to fix it. This means that we exclude those bug-fixes for which the change in a method does not completely fix the bug, ensuring a fair training and evaluation process.
We exclude $\sim$1.5M instances in this step.
Then, following the state-of-the-art pre-processing procedures, we perform additional filtering steps. Specifically, we exclude instances with commented-out code \cite{shi2022we} (106,487), unusual sequences like \texttt{``\_\_\_\_''} \cite{tufano2023automating} (13,245), and self-admitted technical debt, like \texttt{TODO} or \texttt{FIXME} \cite{mastropaolo2021empirical, shi2022we, tufano2023automating} (20,123).
To avoid to confuse the model during fine-tuning, we remove instances for which we have the same ``buggy'' code and different ``fixed'' code, and vice-versa \cite{vitale2023using} (12,971).

Unlike Tufano \etal \cite{tufano2019empirical}, we do not abstract identifiers or literals, preserving the source code in its original form. We exclude instances with fewer than 10 tokens (trivial methods) and those exceeding 512 tokens, as the model cannot handle more than this number of tokens. This filtering approach aligns with similar works \cite{tufano2022using, mastropaolo2022using, vitale2023using}. As a result, our dataset includes larger instances compared to the dataset by Tufano \etal \cite{tufano2019empirical}, which is limited to methods with a maximum of 100 tokens, as shown in \tabref{tab:metrics_datasets}. After applying these filters, we obtain a dataset comprising 116,372 bug-fix pairs, which we refer to as $D_b$.

\begin{table}[t]
\caption{Comparing the old and renewed datasets in terms of number of tokens and cyclomatic complexity (CC) for both the \textit{buggy} (B) and \textit{fixed} (F) parts of the instances.}
\centering
\resizebox{0.9\linewidth}{!}{%
\begin{tabular}{l|r|r|r|r}
\toprule
\textbf{Dataset} & \textbf{\# Tokens (B)} & \textbf{\# Tokens (F)} & \textbf{CC (B)} & \textbf{CC (F)}\\
\midrule
\textit{BFP}$_{small}$  &  31.92 &  29.16 & 1.35 & 1.36 \\
\textit{BFP}$_{medium}$ &  75.07 &  73.68 & 2.45 & 2.50 \\
\textit{Renewed (Our)}  & 101.29 & 108.36 & 3.22 & 3.49 \\
\bottomrule
\end{tabular}
}
\label{tab:metrics_datasets}
\end{table}

\textbf{Automatically Generating Comments.}
To validate our theory, we require two datasets: one containing commented instances and another with the same instances stripped of comments (see the experimental procedure for more details). A potential approach could have been to use only the instances in $D_b$ that include developer-written comments as the dataset with comments, and create the dataset without comments by removing those comments.
However, we observed that only a small fraction of instances in $D_b$ contain developer-written comments (approximately 20\%). Moreover, many of these comments are short or incomplete. For example, a method named \texttt{format}, which formats and appends various types of \texttt{units} to a string, is documented with the one-word comment \texttt{Formatting}. As a result, relying solely on instances with sufficiently detailed developer-written comments would have produced a dataset that is both small and inconsistent.
To address this, we adopted the opposite approach: We retained only the code from the instances in $D_b$ and used an automated methodology to generate comments, creating the alternative dataset with comments through automated comment generation.

As a preliminarily step, we removed all the developer-written comments from the methods in $D_b$. Then, we relied on GPT-3.5 \cite{brown2020language} to generate multi-intent comments for a method to document with an in-context learning setting \cite{brown2020language}, similarly to Geng \etal \cite{geng2024large}. The authors showed that providing examples in the prompts helps the models to generate comments that address the different aspects of the method (\secref{subsec:what_comments_matter}).
We relied on GPT-3.5 to balance the performance of GPT-based models with a manageable cost and because Su and McMillan \cite{su2024distilled} found that it generates code summaries that are preferred by developers even more than those used as 'ground-truth' in existing datasets.
Based on our theory reported in \secref{sec:theory}, we generate comments for such intents (\ie \textit{what}, \textit{how}, \textit{why}, \textit{usage}, and \textit{property}). 
We use a prompt template with 2 examples (two-shot learning) that provides GPT-3.5 with sufficient information to generate the comments. We explicitly ask the model to generate a single sentence comment for each category for two reasons. First, this choice aligns with code-summarization research, in which the objective is, typically, to generate a single sentence summary \cite{ahmad2020transformer,haque2020improved,shi2022we,su2024distilled}. Second, we want to avoid an overly unrealistic scenario in which comments exceed code (which would have happened for smaller methods if we did not impose this constraint).
The prompt template we adopted is reported below:
\newcommand{\PField}[1]{\texttt{\{#1\}}}

\begin{examplebox}
 You are an expert software developer and you are great at writing comments as requested. Your role is to write a single sentence of comment for each one of the following categories: what, why, how-it-is-done, how-to-use, property.
 Below you find some examples: \PField{example-1}; \PField{example-2}.
 Now write the comment for the following Java method: \PField{target-method}.
\end{examplebox}

\begin{figure*}[t]
    \centering
    \includegraphics[width=0.77\linewidth]{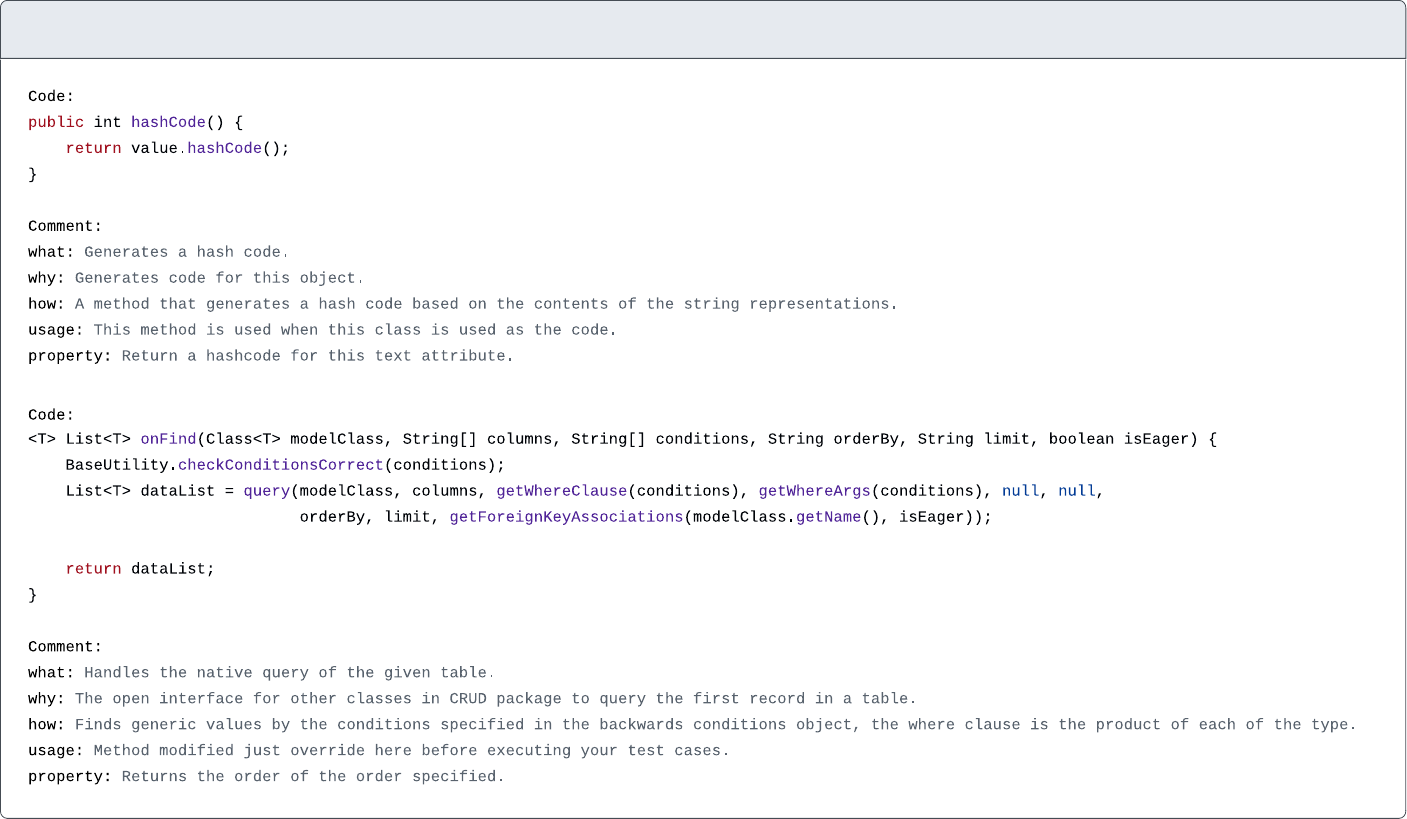}
    \caption{The two examples we use for few-shot learning in our prompt.}
    \label{fig:study:twoshot}
\end{figure*}

The two examples (\PField{example-1} and \PField{example-2}) are reported in \figref{fig:study:twoshot}. We choose the two explicit examples of multi-intent comment generation from the replication package of the work by Mu \etal \cite{mu2023developer}. One of them is a trivial example (first one in \figref{fig:study:twoshot}), while the second one is more complex (second one in \figref{fig:study:twoshot}) to expose the model to different levels of challenge.
The only part of the whole prompt template that changes for each instance is \PField{target-method}.
Given an instance $\langle m^{\mathit{buggy}}_i, m^{\mathit{fixed}}_i \rangle$ in $D_b$, we replace \PField{target-method} with $m^{\mathit{fixed}}_i$, \ie the \textit{fixed} version of the method. 
We generate comments for the \textit{fixed} version rather than for the \textit{buggy} version because the latter code does not completely meet the requirements (by definition). The comments generated from it might reflect such errors. The presence of wrong comments might not allow us to test our theory.
For each instance $\langle m^{\mathit{buggy}}_i, m^{\mathit{fixed}}_i \rangle$ in $D_b$, we define a comment-augmented alternative by pre-pending the generating comments to both the \textit{buggy} and \textit{fixed} versions, leading to $\langle CG(m^{\mathit{fixed}}_i) + m^{\mathit{buggy}}_i, CG(m^{\mathit{fixed}}_i) + m^{\mathit{fixed}}_i \rangle$, where $CG$ is the comment generation methodology we previously defined. This leads to the definition of our comment-augmented alternative of $D_b$, which we call $D_c$.
We did not tune the prompt to optimize the quality of the generated comments since proposing a comment generation approach was beyond the scope of our work.

It is worth noting that using the fixed version of the code to generate comments makes this methodology inapplicable as-is for improving ABF in practical scenarios, as the \textit{fixed} version is obviously not available beforehand. However, our objective is not to propose a practical methodology for enhancing bug-fixing but to validate our theory and determine whether comments play a significant role in the first place.

\textbf{Validating Generated Comments.}
We conducted a manual analysis on a random sample including 400 instances to validate the generated comments. Specifically, we evaluated the \textit{coherence} \cite{mu2023developer}, defined as whether a comment correctly reflects its intended category (\eg what, why, how), and the \textit{adequacy} \cite{mcburney2014automatic}, defined as the extent to which the comment correctly describes those aspects of the source code and does not make mistakes. Coherence is assessed in a binary way (\textit{coherent} or \textit{not coherent}), while adequacy is assessed on a 1 to 5 Likert scale.
Two of the authors independently performed such an analysis and discussed any disagreement trying to reach consensus. As for adequacy, we considered two independent evaluations as not in agreement if one of them was positive (4 or 5) and the other one was negative (1 or 2). In the other cases, we computed the overall adequacy as the mean of the two assigned scores.
In the \textit{coherence} analysis, the evaluators initially agreed on 349 instances (87\% agreement), while the remaining 51 were carefully reviewed and resolved in a meeting. In contrast, no disagreements arose in the \textit{adequacy} analysis, for which we observed a substantial agreement (Cohen's $\kappa$ \cite{cohen1960coefficient} = 0.72).
We observed that 93\% of the analyzed comments are coherent with the related comment category and that the median adequacy score assigned is equal to 4. Therefore, we can claim that the generated comments are sufficiently good for our experiment.

\textbf{Final Filtering and Splitting.}
The comment-augmented dataset $D_c$ naturally contains larger instances than $D_b$ (it contains the same code plus the generated comments). For this reason, some of the instances in $D_c$ contain more than 512 tokens (the maximum our model can handle). Thus, we remove from $D_c$ the instances that contain more than 512 tokens. We remove such instances also from $D_b$, ensuring that we still have the same methods in both the datasets. In this step, we remove 6,163 instances from both datasets.
We also discarded from both the dataset all the instances for which GPT-3.5 was not able to correctly generate the comments (110).
We identified such instances by automatically verifying, via regular expressions, that the generated comments followed the expected format, composed all the types (\eg \texttt{what:}) followed by a sentence.

We found those instances by filtering out those that did not respect the expected comments structure (\ie ``what: {what-comment}'', \etc).
We randomly partition the instances in $D_b$ into three sets (fine-tuning $\Phi_b$, test $T_b$, and validation $V_b$) following the 80-10-10 rule, like most previous work did \cite{vitale2023using,mastropaolo2024toward,mastropaolo2021studying,berabi2021tfix}.
We then split $D_c$ into fine-tuning, test and validation sets ($\Phi_c$, $T_c$, and $V_c$, respectively) by making sure that each instance $i_c \in D_c$ was assigned to the same type of set as $D_b$. For example, if an instance $i_b \in D_b$ was assigned to the fine-tuning set $\Phi_b$, its commented counterpart, $i_c \in D_c$ was assigned to the fine-tuning set $\Phi_c$.
As a result, we reserved 93,098 pairs for the fine-tuning sets ($\Phi_b$ and $\Phi_c$), 11,637 for the test sets ($T_b$ and $T_c$), and 11,637 for for the validation  ($V_b$ and $V_c$). 

\newcommand{\IV}[1]{$\mathit{IV}_{\mathit{#1}}$}
\subsection{Experimental Procedure}
\label{subsec:exp_proc}
\textbf{\RQ{1}: Comments Impact.}
To answer \RQ{1}, we conduct a \textit{controlled experiment} \cite{wohlin2012experimentation}. The experimental \textit{objects} are both (i) the fine-tuned models, and (ii) the test sets with and without comments we previously defined.
We identified several independent variables (IVs) related to the model training and configuration that might affect the dependent variables (DVs), which are related to the effectiveness of the model. We manipulate some of them (the \textit{factors}, \ie the variables for which we want to study the effect on the DVs). We describe below in details the variables and the procedure we used to conduct the experiment.

We use two LLMs in our experiment: CodeT5+ \cite{wang2023codet5+} and DeepSeek-Coder \cite{guo2024deepseek}.
As for CodeT5+, we employ the smallest variant with 220M parameters, using an encoder-decoder architecture suitable for the bug-fixing task \cite{tufano2019empirical, mastropaolo2021studying, berabi2021tfix}. As for DeepSeek-Coder, we employ the smallest instruction variant with 1.3B parameters, which features a decoder-only architecture.
The first independent variable (\IV{LLM}) takes two values: CodeT5+ and DeepSeek-Coder.
We choose to leverage such models since (i) they are specifically pre-trained to handle coding tasks, (ii) they have been extensively used in previous work \cite{wang2023rap,weyssow2023exploring,kim2024datarecipe,cassano2023can}, and (iii) they employ different model architectures (\ie encoder-decoder and decoder-only).
For each LLM, we fine-tune two versions of the model using different datasets: one of them ($\Phi_c$) contains instances with the comments generated with the previously explained procedure (\IV{Comments_{train}} = \textit{Yes}), and the other one ($\Phi_b$) contains the same instances without any comment (\IV{Comments_{train}} = \textit{No}).
Formally, the independent variable \IV{Comments_{train}}, which represents the presence of comments in fine-tuning instances, is the first \textit{factor} in our experiment, for which we have the two \textit{treatments}, \ie \textit{Yes} (\textit{all} the code instances have a comment) and \textit{No} (\textit{none} of the code instances has comments).
We call the models fine-tuned on the datasets $M_{\Phi_c}^{CT}$, $M_{\Phi_b}^{CT}$ (CodeT5+), and $M_{\Phi_c}^{DS}$, $M_{\Phi_b}^{DS}$ (DeepSeek-Coder), respectively.
The fine-tuning process differs between the models. For CodeT5+ the number of epochs is fixed to 50 (\IV{Epochs} = 50), the batch size is 12 (\IV{BS} = 12), the optimizer is AdamW (\IV{opt} = AdamW), and the input and output size is limited to a maximum of 512 tokens (\IV{input} = \IV{output} = 512). For DeepSeek-Coder, the number of epochs is fixed to 10 (\IV{Epochs} = 10), the batch size is 32 (\IV{BS} = 32), and the optimizer is AdamW (\IV{opt} = AdamW).
Differently from encoder-decoder models, decoder-only models processes the input and generated output tokens as a single continuous sequence during generation. For this reason, we fixed a combined sequence length limit of 2048 tokens (\IV{max\_length} = 2048), ensuring that the total number of input and output tokens does not exceed this value.
We employ early stopping (\IV{ES} = \textit{Yes}) to prevent overfitting \cite{prechelt2002early}. Specifically, we evaluate the models after each epoch on the respective validation sets ($V_c$ for $M_{\Phi_c}^{CT}$ and $M_{\Phi_c}^{DS}$, and $V_b$ for $M_{\Phi_b}^{CT}$ and $M_{\Phi_b}^{DS}$) and take the best checkpoint before the accuracy decreases for five consecutive epochs (patience).

We evaluate each fine-tuned model on both types of test sets to study the impact of comments during both fine-tuning and inference.
This corresponds to the independent variable \IV{Comments_{test}}, which controls the presence of comments in the test instances (\textit{Yes}/\textit{No}).
Specifically, $M_{\Phi_b}^{CT}$ and $M_{\Phi_b}^{DS}$ (trained without comments) are tested on both $T_b$ (\IV{Comments_{test}}=\textit{No}) and $T_c$ (\IV{Comments_{test}}=\textit{Yes}), and $M_{\Phi_c}^{CT}$ and $M_{\Phi_c}^{DS}$ (trained with comments) are also tested on both $T_b$ and $T_c$.
We use greedy decoding as decoding strategy, \ie the model generates the token with the highest probability at each generation step. We choose such a strategy rather than using sampling-based decoding methods to study the actual models learned behaviours during the fine-tuning process.

We consider two \textit{dependent variables}: Exact Match (EM) and CodeBLEU (CB). EM represents the number of the correct inferences made by the model on the test set divided by the size of the test set.
CB \cite{ren2020codebleu} measures the similarity between the inference made by the model and the expected generated code. Such a metric is a variation of BLEU \cite{papineni2002bleu} specialized on the source code. While BLEU relies on matching n-grams, CodeBLEU considers the syntax (through the Abstract Syntax Tree) and its semantics (via data-flow).

In addition, we perform statistical tests to compare the different training and inference configurations in terms of EM.
We use McNemar's test \cite{mcnemar1947note} on paired test instances, along with the odds ratio (OR) to measure the magnitude of the effect size.
We reject the null hypothesis (\ie no difference in effectiveness between two configurations) if the $p$-value is lower than 0.05.
To interpret the magnitude of the effects, we categorize odds ratios according to Cohen's guidelines \cite{cohen2013statistical,rlibrary}: \textit{very small} (OR $< 1.44$), \textit{small} ($1.44 \leq$ OR $< 2.48$), \textit{medium} ($2.48 \leq$ OR $< 4.27$), and \textit{large} (OR $\geq 4.27$).

\textbf{\RQ{2}: Comments Importance.}
To answer \RQ{2}, we employ post-hoc interpretability techniques to understand to what extent tokens that are part of the comments are important in the inferences the models $M_{\Phi_c}^{CT}$ and $M_{\Phi_c}^{DS}$ make.
We use SHAP (SHapley Additive exPlanations) \cite{lundberg2017unified}, a black-box model-agnostic method. As previously explained, LLMs (and thus our models) generate one token at a time. Given a token $t_k$ that the model has to generate from the given context $\chi_{t_k}$, related to the instance $i_c \in T_c$, SHAP assigns a \textit{feature importance} value $f$ to each token $t \in \chi_{t_k}$.
For our analyses, we use the same methodology employed by Liu \etal \cite{liu2024reliability}, \ie we adopt GradientExplainer to approximates SHAP values.
More specifically, for each instance of the test set $\langle m^{\mathit{buggy}}_i, m^{\mathit{fixed}}_i \rangle \in T_c$ that the model correctly predicts, we compute the $f$ values of all the tokens in $m^{\mathit{buggy}}_i$ for each token that the model generates. We ignore the $f$ values assigned to the tokens that the model progressively generates since we are interested in the role of the input code. We normalize the SHAP values so that their sum is equal to 1.

In a first analysis, we aimed to understand the importance given to the comments as compared to the importance given to the source code. For each instance $i_c \in T_c$, we partition the tokens in two sets: tokens belonging to the comments ($C_{i_c}$) and tokens belonging to the source code ($S_{i_c}$). We then estimate the importance of tokens in the comments ($I_{\mathit{Comments}}(i_c)$) by summing the $f$-values assigned to the tokens in $C_{i_c}$, and the importance of tokens in the code $I_{\mathit{Code}}(i_c)$ by summing the $f$-values assigned to the tokens in $S_{i_c}$.
We report the distribution of $I_{\mathit{Comments}}(i_c)$ and $I_{\mathit{Code}}(i_c)$ for all the instances.
In a second more fine-grained analysis, we aimed to understand what is the importance assigned to the five categories of comments we generated for each instance. To do this, for each instance $i_c \in T_c$, we focused on the tokens belonging to the comments ($C_{i_c}$) and normalized the $f$-values computed on them so that their sum is equal to 1. We partitioned the tokens in five classes based on the specific comment they belonged to (\textit{what}, \textit{why}, \textit{how}, \textit{usage}, and \textit{property}). We could easily do this because the comments were always generated in the same order with our prompt (\eg the first comment was always aimed at explaining \textit{what} the method does).
We computed the importance for each of them, again, by summing the $f$-values. We report the distribution of $I_{\mathit{What}}(i_c)$, $I_{\mathit{Why}}(i_c)$, $I_{\mathit{How}}(i_c)$, $I_{\mathit{Usage}}(i_c)$, and $I_{\mathit{Property}}(i_c)$ for all the instances.

%% file: results.tex

\section{Empirical Study Results} \label{sec:results}

We report the results of our empirical study and the answers to our two RQs.

\subsection{RQ$_{1}$: Comments Impact}
\begin{table}[t]
\centering
\caption{Comparison between the models trained and tested with and without comments.}
\resizebox{0.8\linewidth}{!}{%
\begin{tabular}{l|cc|cc}
\toprule
\multirow{2}{*}{\textbf{Models}} & \multicolumn{2}{c|}{$\boldsymbol{T_b}$} & \multicolumn{2}{c}{$\boldsymbol{T_c}$} \\ \cline{2-5}
 & \textbf{EM} & \textbf{CB} & \textbf{EM} & \textbf{CB} \\
\midrule
$M_{\Phi_b}^{CT}$ (\textit{Comments} = No)  & 3.26\% & 76.21\% & 3.73\% & 75.66\% \\
$M_{\Phi_c}^{CT}$ (\textit{Comments} = Yes) & 2.78\% & 78.18\% & 9.80\% & 80.38\% \\
\midrule
$M_{\Phi_b}^{DS}$ (\textit{Comments} = No)  & 5.57\% & 77.72\% & 7.90\% & 86.40\% \\
$M_{\Phi_c}^{DS}$ (\textit{Comments} = Yes) & 5.09\% & 77.37\% & 12.77\% & 81.01\% \\
\bottomrule
\end{tabular}%
}
\label{tab:performance_topk}
\end{table}

The results obtained in terms of Exact Match (EM) and CodeBLEU (CB) for the four fine-tuned models—$M_{\Phi_b}^{CT}$ and $M_{\Phi_b}^{DS}$ (trained without comments), and $M_{\Phi_c}^{CT}$ and $M_{\Phi_c}^{DS}$ (trained with comments)—evaluated on the two test sets $T_b$ (without comments) and $T_c$ (with comments) are summarized in \tabref{tab:performance_topk}.
All differences discussed in the following are statistically significant according to McNemar's test ($p$-value $< 0.05$). This is due to the large size of the test set. Therefore, we focus the discussion on the effect size, as measured by the odds ratio (OR).

We start by analyzing the results on the test set $T_b$, where no comments are provided to the model at inference time.
Under this setting, the CodeT5+ model trained without comments ($M_{\Phi_b}^{CT}$) achieves an EM of 3.26\% and a CB of 76.21\%.
When trained with commented instances ($M_{\Phi_c}^{CT}$), the model reaches an EM of 2.78\% and a slightly higher CB of 78.18\%.
This difference, while statistically significant, reports a \textit{very small} effect size (OR = 0.72), indicating a negligible practical impact.
A similar trend is observed for DeepSeek-Coder: the model without comments in training ($M_{\Phi_b}^{DS}$) achieves an EM of 5.57\% and CB of 77.72\%, while the model trained with comments ($M_{\Phi_c}^{DS}$) reports an EM of 5.09\% and CB of 77.37\%.
Such a comparison corresponds to a \textit{very small} effect size (OR = 0.65).
These results indicate that, when comments are absent at test time, models trained with comments achieve slightly lower performance than those trained without them.
Surprisingly, this behavior suggests that the presence of comments during training does not introduce harmful biases but leads to only slightly weaker performance in contexts lacking comments, likely because such models tend to rely on contextual information that are absent at test time.
\vspace{4pt}
\begin{findbox}
Training with comments does not significantly impact performance in comments-free settings.
\end{findbox}


When comments are included at test time ($T_c$), both model families benefit substantially.
For CodeT5+, the non-commented variant ($M_{\Phi_b}^{CT}$) increases its EM from 3.26\% to 3.73\% (OR = 1.44), showing that comments in the input alone can provide marginal guidance even without prior exposure.
However, the most evident result emerges for the model trained with comments ($M_{\Phi_c}^{CT}$), whose EM increase from 2.78\% to 9.80\% (\ie $\sim$3$\times$ higher) and CB from 78.18\% to 80.38\%, with a \textit{large} effect size (OR = 6.86).
A similar, and more pronounced pattern appears for DeepSeek-Coder: $M_{\Phi_b}^{DS}$ improves from 5.57\% to 7.90\% EM when tested on commented code (\textit{medium} effect size, OR = 3.42), while $M_{\Phi_c}^{DS}$ achieves the best overall results, with an EM of 12.77\% and CB of 81.01\%, approximately $\sim$2$\times$ the previous 5.09\% with a \textit{large} effect size (OR = 9.76).

These results highlight two main behaviors.
First, models trained without comments still benefit from comments at inference time, fixing more bugs when contextual information is available.
Second, models trained with comments \textit{significantly} benefit from such contextual information, reporting performance improvements of approximately two to three times when comments are provided at test time.
In both model families, the combination of training and testing with comments results in the largest performance, showing that comments act as useful auxiliary signals that enhance the models reasoning and code comprehension abilities.
\vspace{4pt}

\begin{findbox}
Comments consistently improve model performance, regardless of the training setup. The improvements are more evident when the model is trained to also leverage such information.
\end{findbox}

Across all configurations, DeepSeek-Coder consistently outperforms CodeT5+, likely due to the larger number of parameters, different architecture, and wider pre-training corpus.
This confirms that larger models generally possess stronger bug-fixing capabilities.
Interestingly, the positive effect of adding comments at inference time, when models have not been trained with them, is more evident for the larger DeepSeek-Coder, suggesting that bigger models can better exploit contextual information even without previous exposure.
On the other hand, when fine-tuning is performed also with commented instances, the smaller CodeT5+ benefits the most, showing a three times EM improvement when tested with comments, compared to approximately two times improvements for DeepSeek-Coder.
This indicates that while larger models are naturally more capable, smaller ones can close much of the gap when guided by explicit contextual information such as comments.
\vspace{4pt}

\begin{rqbox}
\textbf{Answer to \RQ{1}.} Comments enhance LLM bug-fixing by providing contextual information that improves performance even when available only at inference time. Training with comments remains neutral in comment-free cases but highly beneficial when comments are present during inference.
\end{rqbox}

\subsection{RQ$_{2}$: Comments Importance}
\begin{figure}[t]
	\centering
	\includegraphics[width=0.9\linewidth]{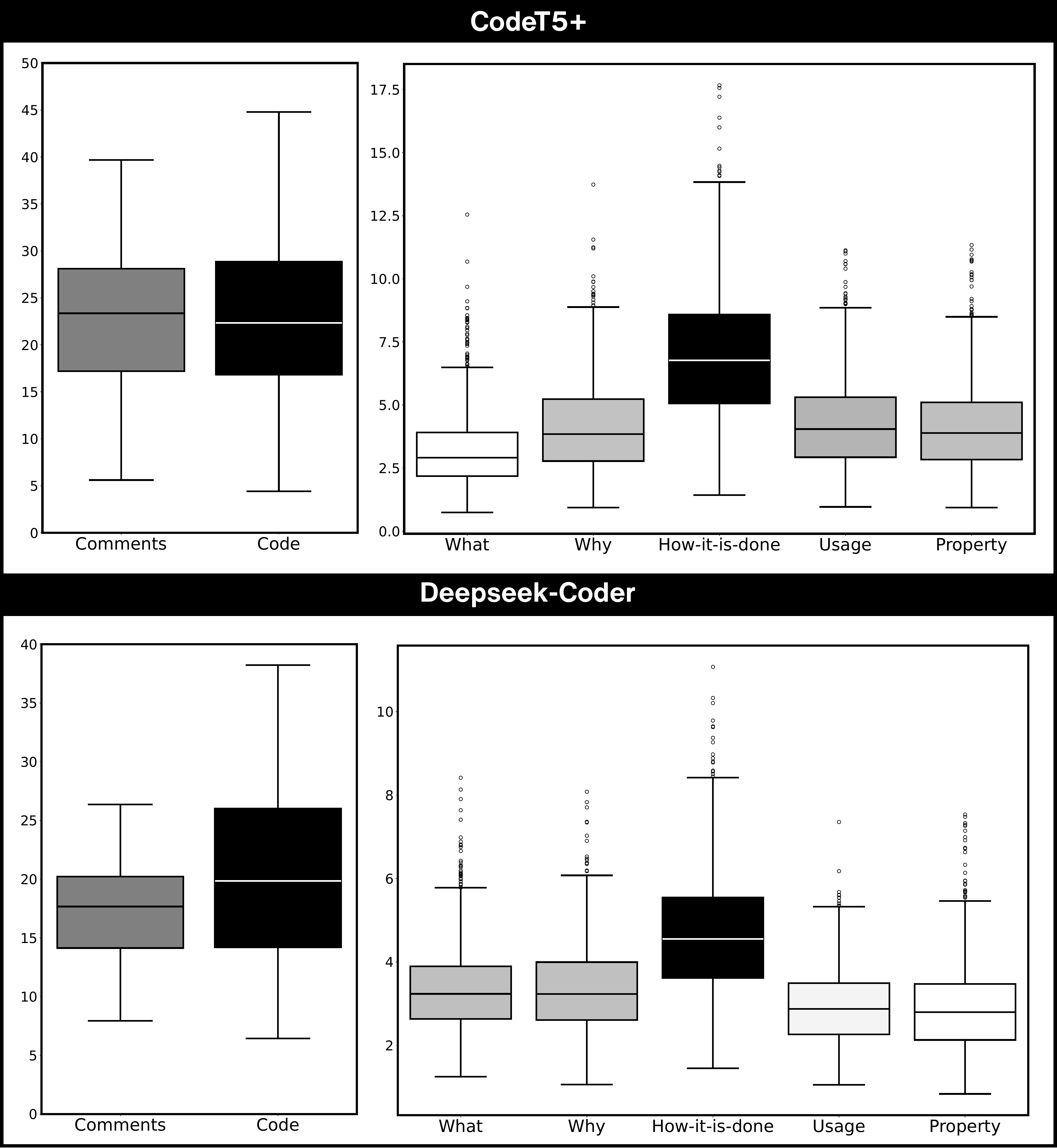}
	\caption{Importance distribution for comments and code (left part) and comment intents (right part).}
	\label{fig:boxplot_comments_code}
\end{figure}

\figref{fig:boxplot_comments_code} illustrates the importance attributed by the models $M_{\Phi_c}^{CT}$ and $M_{\Phi_c}^{DS}$ to tokens belonging to code and comments for their respective 1,141 and 1,486 exact matches.
We find that the models consistently assign higher importance to code-related tokens (\ie those representing the logic of the method) in nearly all instances, with fewer than 100 exceptions for both models. However, the boxplot distributions reveal distinct patterns: for $M_{\Phi_c}^{CT}$, the importance is nearly evenly distributed between code and comments, whereas $M_{\Phi_c}^{DS}$ assigns greater importance to code tokens while still recognizing the value of comment tokens. This analysis, from a different perspective, supports the findings from \RQ{1}: Comments contribute significantly to the fix generation process.

When we focus on the different comment categories (right part of \figref{fig:boxplot_comments_code}), we find that some comments are more important than others.
Both models give much higher importance to comments describing \textit{how} the implementation is done as compared to the other categories.
This is quite expected: Such comments document some implementation details that might have been violated in the \textit{buggy} version.
Thus, the models are able to find the violation and fix it thanks to them.
Then, we find that the $M_{\Phi_c}^{CT}$ gives similar importance to the \textit{property}-related comments, the ones describing the method \textit{usage}, and its design rationale (\textit{why}). On the other hand, it gives less importance to the comments aimed at describing \textit{what} the method does. We conjecture that this happens for two reasons. First, there are generally other elements that help the model grasp \textit{what} the method is aimed to do (\eg the method name or the identifiers). Such elements make such a comment less relevant than others. Second, it is probably more rare that a bug makes the method do something different from what is supposed to do, so that a comment describing what the method does helps the model fixing it. Regarding $M_{\Phi_c}^{DS}$, we find that the model gives lower importance attributions to \textit{property}-related and \textit{usage}-related comments, while giving higher importance to \textit{why} and \textit{what} comments categories.
The models give contrasting importance to \textit{what} comments. We conjecture that the different pre-training objectives employed between the models may impact on such findings \cite{wang2023codet5+,guo2024deepseek}. In particular, the base DeepSeek-coder model was pre-trained using a next-token prediction objective. For this reason, instances used during pre-training might have included inputs that better align with the high-level descriptions generally found in \textit{what} comments.
\vspace{4pt}
\begin{rqbox}
\textbf{Answer to \RQ{2}.} Code comments are assigned significant importance by the models. The most important comments are those documenting implementation details, while the least important are \textit{what} comments for $M_{\Phi_c}^{CT}$ and \textit{property} comments for $M_{\Phi_c}^{DS}$.
\end{rqbox}

%% file: discussion.tex
\section{Discussions and Implications}
\label{sec:discussion}

\begin{figure*}[t]
    \centering
    \includegraphics[width=0.9\textwidth]{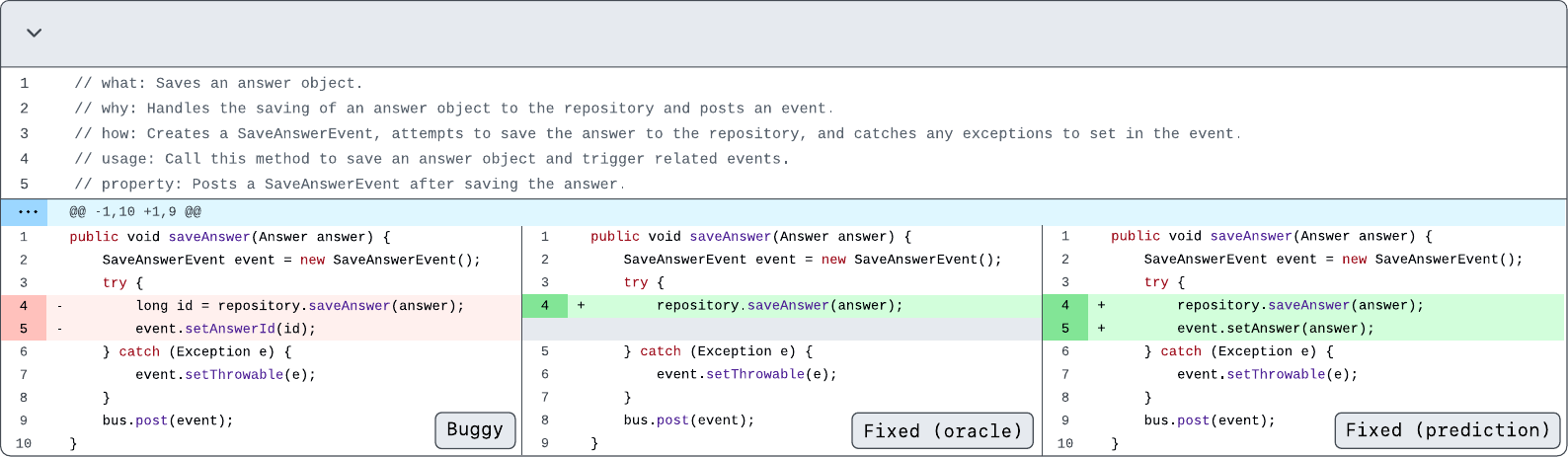}
    \caption{Example of wrong prediction.}
    \label{fig:comm_confuse}
\end{figure*}

\begin{figure}[t]
    \centering
    \includegraphics[width=\linewidth]{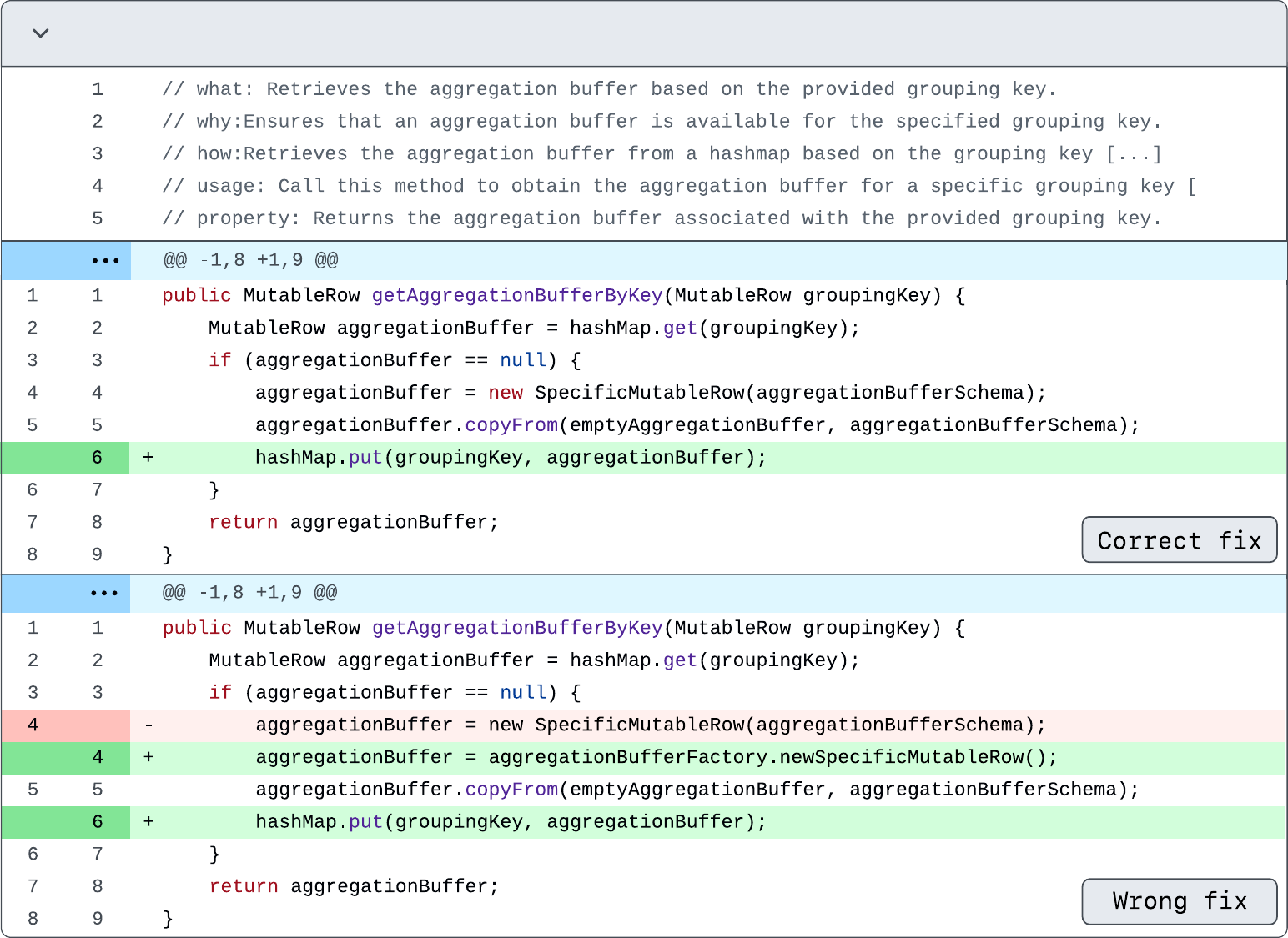}
    \caption{Example of a $M_{\Phi_c}^{CT}$ correct fix, and $M_{\Phi_b}^{CT}$ wrong fix. }
    \label{fig:comm_helps}
\end{figure}
In this section, we discuss the lessons learned and implications for future research based on the results of our controlled experiment.
\subsection{Lessons Learned}
Our results clearly show that comments are beneficial to improve the effectiveness of LLMs in the ABF task.
While this is true in most cases, we found some instances for the generated comments are not sufficiently detailed to fix the bug. In \figref{fig:comm_confuse} we see an interesting example. In the buggy version, the method attempted to set an answer \texttt{``id''} in the \texttt{``SaveAnswerEvent''} object, which could result in incomplete event data if an exception occurred during the save operation.
The fix removed such an assignment, ensuring the event is posted without depending on the successful retrieval of the \texttt{id}. 
However, the model predicted a candidate fix that set the \texttt{``answer''} object in the event, leading to an analogous problem. 
It probably did that because the input code contained a call to the \texttt{setAnswerId} method and the model did not try to change much the code structure. 
In addition, the \textit{how} comment states \textit{``...  and catches any exceptions to set in the event.''}, which could mislead the model. 
We tried to change the comment by removing such a part of the comment, and this resulted in a correct fix.

On the other hand, there are cases in which generated comments are sufficient to help the model generate the correct fix. As we can see in \figref{fig:comm_helps}, the method did not store the \texttt{aggregationBuffer} in the \texttt{hashMap}. The \textit{property}-related comment most likely helped fix this bug since it explicitly says that the method \textit{ensures that an aggregation buffer is available for the specified grouping key}, property that is violated in the buggy version of the code. The model without comments fails to correctly fix the method by using \texttt{aggregationBufferFactory} instead of storing the created \texttt{SpecificMutableRow} in the \texttt{hashMap}.
\vspace{4pt}

\begin{findbox}
\textbf{Lesson Learned 1.} Although comments can boost the performance of automated bug-fixing models, their completeness and correctness is key to fix bugs.
\end{findbox}

In a real use-case scenario, however, it is unlikely that developers provide comments for all the categories we studied. Most importantly, developers might be reticent to provide comments regarding the implementation details because the such details might change frequently, thus leading to inconsistencies between comments and code and an increased effort. 
All the other categories of comments, instead, might change less frequently.
To test what would happen in a more realistic context, we fine-tuned and evaluated $M_{\Phi_c*}^{CT}$ on a variation of the $D_c$ dataset ($D_{c*}$), from which we removed the \textit{how} comments, and tested it on the corresponding test set $T_{c*}$, where the \textit{how} comments were also removed.
Even without implementation details, we still observe a significant boost in terms of EM, which is 91\% higher than the baseline model ($M_{\Phi_b}^{CT}$ on $T_b$), with higher CodeBLEU.
This result shows that even if not too detailed, comments can help improving ABF. 
\vspace{4pt}

\begin{findbox}
\textbf{Lesson Learned 2.} Even non-implementation related comments can help to improve the effectiveness of LLMs for ABF.
\end{findbox}

We further explored what would happen if the comments were instead generated from the \textit{buggy} version of the code.
This scenario represents a realistic context, in which comments can be automatically generated or manually written based on the available buggy code.
However, this setting \textit{inherently} implies incorrect comments, since they document the buggy logic rather than the intended correct behavior.
Thus, we tested the previously fine-tuned models on a variant of the test set with comments generated from the buggy code.
As expected, all models performed the same or slightly worse with buggy-generated comments. For instance, CodeT5+ saw small EM drops (3.26\% to 2.92\%, 2.78\% to 2.67\%), with similar negligible or negative trends for DeepSeek-Coder.
These results confirm that comments generated from buggy code do not provide useful guidance and can even hinder models, since they reinforce incorrect behaviors embedded in the buggy code.
\vspace{4pt}

\begin{findbox}
\textbf{Lesson Learned 3.} Comments generated from buggy code do not help and can mislead automated bug-fixing models.
\end{findbox}
\vspace{-10pt}

\subsection{Implications}
\label{disc:implications}
Our findings have several implication for both researchers and practitioners.
First, removing code comments from training instances is a common practice in ABF research \cite{tufano2019empirical}. However, given the results of our findings we suggest researchers to not remove comments from training instances. Low-quality comments should be improved, whenever possible. Moreover, researchers could devise approaches to automatically generate high-quality comments that serve as effective contextual information, inferred from the broader project context.
For practitioners, our findings highlight the importance of thoroughly documenting code to better leverage the capabilities of LLMs for automating bug-fixing. While it is widely recognized that code documentation improves maintainability, our results suggest that its benefits extend beyond internal quality, positively impacting external quality by enhancing the support developers receive in the bug fixing process by LLMs. Although our study focused on ABF, we conjecture that these findings may apply to other tasks as well, which future research should aim to confirm or refute.
Finally, tool builders could develop features to identify comments lacking sufficient detail (\eg those that fail to explain rationale, preconditions, or postconditions) or of low quality (\eg inconsistent with the code). Such tools could alert developers when attempting to use LLMs to repair methods with inadequate comments and encourage improvements, potentially supported by automated suggestions.

%% file: threats.tex
\section{Threats to Validity} \label{sec:threats}
\textbf{Threats to construct validity}. We started from the dataset provided by Tufano \etal \cite{tufano2019empirical} to define the datasets we employ in our study, \ie $D_b$ and $D_c$. As the authors themselves acknowledged, some instances might not represent actual bug-fixing instances or the bug-fixing change might be in a method different from the one in a given instance. To limit this threat, we used a stricter criterion to select instances, \ie we only considered commits impacting a single method.
We defined $D_c$ leveraging GPT-3.5 to generate the multi-intent comments. It is possible that some generated comments are not appropriate for a given category (\eg \textit{usage}). Besides, it is possible that such comments are not good enough. To limit this threat, we conducted a manual analysis on random sample including 400 generated comments. We evaluated the \textit{coherence} to the given category (\textit{coherent} or \textit{not coherent}) and the \textit{adequacy} to the source code (on a 1 to 5 Likert scale). Two of the authors independently performed such an analysis. If any disagreement arose, they would have discussed trying to reach a consensus.
It is also possible that some instances in our dataset appeared in the datasets used to pre-train CodeT5+ and DeepSeek-Coder. We started from the same pre-trained models to define the specialized bug-fixing models trained with and without comments.
If such an issue biased the results, it did so for both the models. Thus, we believe that the \textit{relative} results (\ie the difference between the models) was not affected by this possible issue whatsoever.

\textbf{Threats to internal validity}. 
We used EM and CB as dependent variables of our experiment to measure the effectiveness of the models. 
EM is a lower bound of the actual bug-fixing capabilities of the models we tested. A prediction that exactly matches the expected output is surely a bug fix; on the other hand, a prediction not exactly matching the excepted output is not necessarily a wrong fix as it could contain a valid alternative patch.
For example, if the expected fix contains a call to \texttt{Collections.isEmpty(list)} while the predicted one contains \texttt{list.isEmpty()}, the latter is a valid bug fix even if not exactly matching the expected one.
Thus, it is possible that the \textit{absolute} capabilities of the models are actually higher.
We use CodeBLEU to measure the distance between the predicted sequence and the expected one. We used this as an estimate of the effort a developer would make to transform an incorrect fix provided by the model into the correct fix. It is worth noting, however, that a small number of tokens to change does not necessarily imply that developers would make a smaller effort.
Finally, we adopted a state-of-the-art post-hoc interpretability methods \cite{lundberg2017unified, liu2024reliability} to estimate the importance given by the model to comments (in general) and to different categories of comments (specifically).
It is worth noting that such an approach, like others available, only provide estimates on the importance given by the model to single tokens. Thus, it is possible that the importance that the model assigns to comments and to different comment categories differs from the one we report. It is worth noting, however, that when removing the supposedly best category of comments according to our analysis (\ie \textit{how} comments) we obtain lower EM than the one obtained when using all the comments (even if still higher than the one obtained without comments). While this suggest that the reported values are sufficiently reliable, further tests with single categories of comments are required to confirm this result.

\textbf{Threats to external validity}.
The dataset by Tufano \etal \cite{tufano2019empirical} from which we built our dataset contains instances relatively old and limited to the Java programming language. Our results might not generalize to bug-fixing commits performed more recently and on different programming languages. It is worth noting that the dataset by Tufano \etal has been used in recent work as well \cite{mastropaolo2022usingt5,lin2024one,hossain2024deep,zirak2024improving, wang2023rap,mastropaolo2021studying,chakraborty2021multi,drain2021generating,wang2021codet5,ahmad2021unified,guo2020graphcodebert,huang2023empirical,huang2025comprehensive}.
Our results might not generalize to other models.
To limit this threat, we tested our theory on two models, CodeT5+ \cite{wang2023codet5+} and DeepSeek-Coder \cite{guo2024deepseek}. The first representative of an encoder-decoder architecture and limited number of parameters (\ie 220M), and the second one representative of a decoder-only architecture with an higher number of parameters (\ie 1.3B).
To further assess the generalizability of our findings, we evaluated GPT-4.1 in a zero-shot setting on the same test sets with and without comments, observing an increase in EM from 6.11\% (without comments) to 8.63\% (with comments). This supplementary result provides preliminary evidence that the positive impact of comments on automated bug fixing extends to more recent and larger LLMs.
Still, we acknowledge that further empirical evidence is needed to confirm this phenomenon on different models and different configurations.

%% file: conclusion.tex
\section{Conclusion} \label{sec:conclusion}
We conducted an empirical study to investigate the impact of comments on the Automated Bug Fixing (ABF) task.
We compared models trained and tested on all combinations of bug-fix pairs with and without automatically generated, multi-intent comments focusing on five intents (\textit{what}, \textit{why}, \textit{how}, \textit{usage}, and \textit{property}).

Our results show that the presence of comments—whether provided during training, testing, or both—substantially improves the effectiveness of large language models in fixing bugs, with accuracy improvements of up to three times compared to configurations where comments are absent.
When analyzing the importance of the five categories of comments we took into account, we found that comments that explain \textit{how} the method is implemented are the most important ones, while the ones related to \textit{what} and/or \textit{property} the method does are less relevant.

Future studies are needed to provide further empirical evidence in support of our findings, also on other tasks (\eg vulnerability fixing). Besides, it will be interesting to study to what extent the comments provided by developers are sufficient to improve the effectiveness in bug fixing.